\def\rfr#1{eq. (\ref{#1})}

\def\Rfr#1{Eq. (\ref{#1})}

\def\asec{$''$ cy$^{-1}$}

\def\dert#1#2{\frac{{{d}}{#1}}{{{d}}{#2}}}              

\def\asec{$''$ cy$^{-1}$}

\def\bar{\begin{eqnarray}}
\def\ear{\end{eqnarray}}
\def\bb{\bibitem}
\def\eqi{\begin{equation}}
\def\eqf{\end{equation}}
\def\eqia{\begin{eqnarray}}
\def\eqfa{\end{eqnarray}}
\def\rp#1#2{{#1\over#2}}

\def\lb#1{\label{#1}}

\def\oc2{$\mathcal{O}(c^{-2})$}

\def\bds#1{\boldsymbol{#1}}

\documentclass{aastex}
\usepackage{url}
\usepackage{amsmath,amsthm,amscd,amssymb}
\usepackage{latexsym}
\usepackage{graphicx,epsfig}

\RequirePackage{color}

\begin{document}

\title{Solar System tests of some models of modified gravity proposed to explain galactic rotation curves without dark matter}

\shorttitle{Solar System tests  of long-range modified gravity models}
\shortauthors{L. Iorio and M.L. Ruggiero}

\author{Lorenzo Iorio }
\affil{INFN-Sezione di Pisa. Permanent address for correspondence: Viale Unit\`{a} di Italia 68, 70125, Bari (BA), Italy. E-mail: lorenzo.iorio@libero.it}
\and
\author{Matteo Luca Ruggiero}
\affil{Dipartimento di Fisica del Politecnico di Torino and INFN-Sezione di Torino Corso Duca degli Abruzzi 24, 10129,
Torino (TO), Italy. E-mail: ruggierom@polito.it}

\begin{abstract}
We consider the recently estimated corrections $\Delta\dot\varpi$ to the Newtonian/Einsteinian secular precessions of the
longitudes of perihelia $\varpi$ of several planets of the Solar System  in order to evaluate  whether they are compatible with the predicted precessions due to models of long-range modified gravity put forth to account for certain features of the rotation curves of galaxies without resorting to dark matter. In particular, we consider  a logarithmic-type correction and a
$f(R)$ inspired  power-law modification of the Newtonian gravitational potential. The results obtained by taking the ratio of the apsidal rates for different pairs of planets show that the modifications of the Newtonian potentials examined in this paper are not compatible with the secular
extra-precessions of the perihelia of the Solar System's planets estimated by E.V. Pitjeva as solve-for parameters processing almost one century of data with the latest EPM ephemerides.
\end{abstract}

\keywords{Experimental tests of gravitational theories; Modified theories of gravity;  Celestial mechanics;  Orbit determination and improvement; Ephemerides, almanacs, and calendars}


\section{Introduction}\lb{intro}

Dark matter and dark energy are, possibly, the most severe
theoretical issues that modern astrophysics and cosmology have to
face, since the available observations seem to  question  the
model of gravitational interaction on a scale larger than the
Solar System. In fact, the data coming from the galactic rotation
curves of spiral galaxies \citep{Binney87}  cannot be explained on
the basis of Newtonian gravity or General Relativity (GR)
if one does not introduce invisible dark matter compensating for the observed inconsistency of the Newtonian dynamics of stars.
Likewise, since 1933, when \citet{Zwi33} studied the velocity dispersion
in the Coma cluster, there is the common agreement on the fact
that the dynamics of cluster of galaxies is poorly understood on
the basis of Newtonian gravity or GR \citep{Peebles93,Peacock99} if the dark matter is not taken into account.
In order to reconcile theoretical models with observations, the
existence of a peculiar form of matter is postulated, the so
called \textit{dark matter}, which is supposed to be a cold and
pressureless medium, whose distribution is that of a spherical
halo around the galaxies.  On the other hand, a lot of
observations, such as the light curves of the type Ia supernovae
and the cosmic microwave background (CMB) experiments
\citep{Riess98,Perlmutter99,Tonry03,Bennet03}, firmly state that
our Universe is now undergoing a phase of accelerated expansion.
Actually, the present acceleration of the Universe cannot be
explained, within GR, unless the existence of
a cosmic fluid having exotic equation of state is postulated:  the so-called
\textit{dark energy}.

The main problem dark matter and dark energy bring about is
understanding their nature, since they are introduced as
\textit{ad hoc} gravity sources in a well defined model of
gravity, i.e. GR (or Newtonian gravity). Of course, another
possibility exists: GR (and its approximation, Newtonian gravity)
is unfit to deal  with gravitational interaction at galactic,
intergalactic and cosmological scales. The latter viewpoint, led
to the introduction of various modifications of gravity \citep{Noj,Espo08}.

However, we must remember that, even though there are problems
with galaxies and clusters dynamics and cosmological observations,
GR is in excellent agreement with the Solar System experiments
(see \citet{Will06} and \citet{Dam06}): hence, every theory that aims at explaining the
large scale dynamics and the accelerated expansion of the Universe,
should reproduce GR at the Solar System scale, i.e. in a suitable
weak field limit. So, the viability of these modified theories of
gravity at Solar System scale  should be studied with great care.

In this paper we wish to quantitatively deal with such a problem by means of the secular (i.e. averaged over one orbital revolution) precessions of the longitudes of the perihelia $\dot\varpi$ of some planets of the Solar System (see also \citep{Iorio07a,Iorio07b}) in the following way.
Generally speaking,
let ${\rm LRMOG}$ (Long-Range Modified Model of Gravity) be a given exotic model of modified gravity parameterized in terms of, say, $K$, in a such a way that $K=0$ would imply no modifications of gravity at all. Let ${\mathcal{P}}({\rm LRMOG})$ be the prediction of a certain effect induced by such a model like, e.g., the secular precession of the perihelion of a planet. For all the exotic models considered it turns out that \eqi{\mathcal{P}}({\rm LRMOG}) = Kg(a,e),\eqf where $g$ is a function of the system's orbital parameters $a$ (semimajor axis) and $e$ (eccentricity); such $g$ is a peculiar consequence of the model ${\rm LRMOG}$ (and of all other models of its class with the same spatial variability).
Now, let us take the ratio of ${\mathcal{P}}({\rm LRMOG})$ for two different systems {\rm A} and {\rm B}, e.g. two Solar System's planets: ${\mathcal{P}}_{\rm A}({\rm LRMOG})/{\mathcal{P}}_{\rm B}({\rm LRMOG}) = g_{\rm A}/g_{\rm B}$. The model's parameter $K$ is now disappeared, but we still have a prediction that retains a peculiar signature of that model, i.e. $g_{\rm A}/g_{\rm B}$. Of course, such a prediction is valid if we assume $K$ is not zero, which is just the case both theoretically (${\rm LRMOG}$ is such that should $K$ be zero, no modifications of gravity at all occurred) and observationally because $K$ is usually determined by other independent long-range astrophysical/cosmological observations. Otherwise, one would have the meaningless prediction $0/0$. The case $K=0$ (or $K\leq\overline{K}$) can be, instead, usually tested  by taking one perihelion precession at a time, as already done, e.g., by \citet{Iorio07a,Iorio07b}.

If we have observational determinations ${\mathcal{O}}$ for {\rm A} and {\rm B} of the effect considered above  such that they are affected  also\footnote{If they are differential quantities constructed by contrasting observations to predictions obtained by analytical force models of canonical effects, ${\mathcal{O}}$ are, in principle, affected also by the mismodelling in them.} by  ${\rm LRMOG}$ (it is just the case for the purely phenomenologically estimated corrections to the standard Newton-Einstein perihelion precessions, since ${\rm LRMOG}$ has not been included in the  dynamical force models of the ephemerides adjusted to the planetary data in the least-square parameters' estimation process by Pitjeva \citep{Pit05a,Pit05b,PitKBO}), we can construct ${\mathcal{O}}_{\rm A}/\mathcal{O}_{\rm B}$ and compare it with the prediction for it by ${\rm LRMOG}$, i.e. with $g_{\rm A}/g_{\rm B}$. Note that $\delta{\mathcal{O}}/{\mathcal{O}}>1$ only means that ${\mathcal{O}}$ is compatible with zero, being possible a nonzero value smaller than $\delta{\mathcal{O}}$. Thus, it is perfectly meaningful to construct ${\mathcal{O}}_{\rm A}/\mathcal{O}_{\rm B}$. Its uncertainty will be conservatively evaluated as $|1/{\mathcal{O}}_{\rm B}|\delta{\mathcal{O}}_{\rm A} + |{\mathcal{O}}_{\rm A}/{\mathcal{O}}_{\rm B}^2|\delta{\mathcal{O}}_{\rm B}$. As a result, ${\mathcal{O}}_{\rm A}/\mathcal{O}_{\rm B}$ will be compatible with zero. Now, the question is: Is it the same for $g_{\rm A}/g_{\rm B}$ as well? If yes, i.e. if \eqi\rp{{\mathcal{O}}_{\rm A}}{\mathcal{O}_{\rm B}}=\rp{{\mathcal{P}}_{\rm A}({\rm LRMOG})}{{\mathcal{P}}_{\rm B}({\rm LRMOG})}\eqf within the errors, or, equivalently, if \eqi\left|\rp{{\mathcal{O}}_{\rm A}}{\mathcal{O}_{\rm B}} - \rp{{\mathcal{P}}_{\rm A}({\rm LRMOG})}{{\mathcal{P}}_{\rm B}({\rm LRMOG})}\right|=0\eqf within the errors,  ${\rm LRMOG}$ survives (and the use of the single perihelion precessions can be used to put upper bounds on $K$). Otherwise, ${\rm LRMOG}$ is ruled out.

The paper is organized as follows. In Section \ref{periprec}
we apply this approach to a test particle in motion around a
central mass $M$ whose Newtonian gravitational potential exhibits
a logarithmic-type correction. Then, in Section \ref{Capozzia},
we consider the power-law modification of the gravitational potential inspired by $f(R)$ extended theories of gravity. Comments and conclusions are outlined in Section \ref{conc}.

\section{The perihelion precession due to a $1/r$ force}\lb{periprec}

Logarithmic corrections to the Newtonian gravitational potential
have been recently used to phenomenologically tackle the problem
of dark matter in galaxies
\citep{vMoo87,Cre00,Kin01,Kir06,Fab07,Sob07}. E.g., \citet{Fab07}
used \eqi V_{\rm ln}=-\alpha GM \ln
\left(\rp{r}{r_0}\right),\lb{pot}\eqf where $\alpha$ has the
dimension of L$^{-1}$, to  satisfactorily fit the rotation curves
of 10 spiral galaxies getting \eqi \alpha\approx -0.1\ {\rm
kpc}^{-1}.\lb{tors}\eqf  The extra-potential of \rfr{pot} yields
an additional $1/r$ radial force\footnote{For another example of a
$1/r$ extra-force and its connection with galaxy rotation curves
see \citep{San06}.}
 \eqi {\bds{A}}_{\rm ln}=\rp{\alpha GM}{r} \ \bds{\hat{r}}.\lb{accel}\eqf
 Various theoretical justifications have been found for a logarithmic-like extra-potential. E.g., according to \citet{Kir06}, it would arise from large-scale discrepancies of the topology of the actual Universe from the Fridman space; \citet{Sob07} obtained it in the framework of the $f(R)$ modifications of general relativity getting preliminarily flat rotation curves, the Tully-Fisher
relation (admittedly with some reservations) and a version of
MOND, while \citet{Fab07} pointed out that string-like objects
\citep{Sol95,Cap06} would yield a logarithmic-type potential with
$\alpha$ related to the string tension.

Alternative tests of such proposed correction to the Newtonian
potential, independent of the dark matter issues themselves, would
be, of course, highly desirable and could be, in principle,
conducted in the Solar System. This is, indeed, considered one of
the tasks to be implemented in further investigations by
\citet{Sob07}; \citet{Fab07} argue that, in view of their extreme
smallness due to \rfr{tors}, no detectable effects induced by
\rfr{accel} would be possible at the level of Solar System.\\

E.V. Pitjeva  has recently processed
almost one century of data of different types for the major bodies of the Solar System in the effort of
continuously improving the EPM2004/EPM2006 planetary ephemerides
\citep{Pit05a,PitKBO}. Among other things, she also simultaneously estimated corrections  $\Delta\dot\varpi$
to the secular rates of the longitudes of perihelia $\varpi$ of the inner
\citep{Pit05b} and of some of the outer \citep{Pit06,PitKBO} planets of
the Solar System as fit-for parameters of a global solution in
which she contrasted, in a least-square way, the observations to
their predicted values computed with a complete set of dynamical
force models including all the known Newtonian and Einsteinian
features of motion. As a consequence, any un-modelled exotic force
present in nature is, in principle, entirely accounted for by the obtained
apsidal extra-precessions\footnote{Of course, since modelling is not perfect, in principle, $\Delta\dot\varpi$ include also the mismodelled parts of the standard Newtonian/Einsteinian effects, but it turns out that their inclusion in the following computation do not alter our conclusions.} $\Delta\dot\varpi$.  See Table \ref{tavola1}
for the inner planets and Table \ref{tavola2} for the gaseous
giant ones.
\begin{table*}
\small
\caption{ Inner planets. First row: estimated perihelion
extra-precessions, from Table 3 of \citep{Pit05b}. The quoted
errors are not the formal ones but are  realistic. The units are
arcseconds per century (\asec). Second row: semimajor axes, in Astronomical Units (AU). Their formal errors are in Table IV of \citep{Pit05a}, in m. Third row: eccentricities. Fourth row: orbital periods in years.\label{tavola1} }

\begin{tabular}{@{}llll@{}}
\hline
& Mercury & Earth & Mars\\
\tableline
$\left<\Delta\dot\varpi\right>$ (\asec) & $-0.0036\pm 0.0050$ & $-0.0002\pm 0.0004$ & $0.0001\pm 0.0005$\\
$a$ (AU) & 0.387 & 1.000 & 1.523 \\
$e$ & 0.2056 & 0.0167 & 0.0934\\
$P$ (yr) & 0.24 &  1.00 & 1.88\\
\hline

\end{tabular}

\end{table*}
\small\begin{table*}
\small
\caption{ Outer planets. First row: estimated perihelion extra-precessions
\citep{Pit06,PitKBO}. The quoted uncertainties are the formal,
statistical errors re-scaled by a factor 10 in order to get the
realistic ones.  The units are
arcseconds per century (\asec). Second row: semimajor axes, in Astronomical Units (AU). Their formal errors are in Table IV of \citep{Pit05a}, in m. Third row: eccentricities. Fourth row: orbital periods in years.\label{tavola2}}

\begin{tabular}{@{}clll@{}}
\hline

& Jupiter & Saturn & Uranus\\
\tableline
 $\left<\Delta\dot\varpi\right>$ (\asec) & $0.0062\pm 0.036$ & $-0.92\pm 2.9$ & $0.57\pm 13.0$\\
$a$ (AU) & 5.203 & 9.537 & 19.191\\
$e$ &  0.0483 & 0.0541 & 0.0471 \\
$P$ (yr) & 11.86 & 29.45 & 84.07\\
\hline

\end{tabular}

\end{table*} %
In regard to them, we must note that modern data sets cover at
least one full orbital revolution only for Jupiter, Saturn and,
barely, Uranus; this is why no secular extra-precessions of
perihelia for Neptune and Pluto are today
available.

In order to make a direct comparison with them, we will now
analytically work out the secular effects induced by the extra-acceleration of
\rfr{accel} on the pericentre of a test particle. To this aim, we
will treat \rfr{accel} as a small perturbation of the Newtonian
monopole. The Gauss equation for the variation of $\varpi$ under
the action of an entirely radial perturbing acceleration $A_r$ is
\eqi\dert\varpi t =-\rp{\sqrt{1-e^2}}{nae}A_r\cos f,\lb{gaus}\eqf
in which $a$ is the semimajor axis, $e$ is the eccentricity,
$n=\sqrt{GM/a^3}$ is the (unperturbed) Keplerian mean motion related to the orbital period $P$ by $n=2\pi/P$, and
$f$ is the true anomaly. After being evaluated onto the
unperturbed Keplerian ellipse, \rfr{accel} must be inserted into
\rfr{gaus}; then, the average over one orbital period $P$
must be performed. It is useful to use the eccentric anomaly $E$
by means of the relations
\begin{equation}\left\{\begin{array}{lll}
r=a(1-e\cos E),\\\\
dt=\rp{(1-e\cos E)}{n}dE
,\\\\
\cos f=\rp{\cos E-e}{1-e\cos E
},\\\\
\sin f=\rp{\sin E\sqrt{1-e^2}}{1-e\cos E}.
\lb{eccen}\end{array}\right.\end{equation}
On using \eqi \rp{1}{2\pi}\int_0^{2\pi}\left(\rp{\cos
E-e}{1-e\cos E}\right)dE = \rp{-1+\sqrt{1-e^2}}{e},\eqf it is
possible to obtain \eqi\left<\dot\varpi\right>_{\rm ln}=-\alpha\sqrt{\rp{GM(1-e^2)}{a}}\left(\rp{-1+\sqrt{1-e^2}}{e^2}\right).\lb{pippo}\eqf
Note that \rfr{pippo} is an exact result with respect to $e$.
 \Rfr{pippo} agrees with the precession obtainable dividing by $P$ the adimensional perihelion shift per orbit $\Delta\theta_p(\rm log)$  worked out by \citet{Adk07} within a different perturbative framework; note that for \citet{Adk07} the constant $\alpha$ has the dimensions of\footnote{Indeed, $V(r)$ of (47) in \citep{Adk07} is an additional potential energy.} M L$^2$ T$^{-2}$, so that that the substitution $\alpha/m\rightarrow -GM\alpha$ must be performed in (50) of   \citep{Adk07} to retrieve \rfr{pippo}.

It may be interesting to note that for the potential of \rfr{pot} the rates for the semimajor
axis and the eccentricity turn out to be zero; it is not so for
the mean anomaly $\mathcal{M}$, but no observational
determinations exist for its extra-rate.

\subsection{Comparison with data}\lb{compa}
According to the general outline of Section \ref{intro},
we may now consider a pair of planets A and B, take the ratio of their estimated extra-rates of perihelia  and compare it to
the prediction of \rfr{pippo} in order to see if they are equal within the errors
 \eqi\Psi_{\rm AB}=\left| \rp{\Delta\dot\varpi_{\rm A}}{ \Delta\dot\varpi_{\rm B} }- \sqrt{ \rp{ a_{\rm B}\left(1-e^2_{\rm A}\right) }{ a_{\rm A}\left(1-e^2_{\rm B}\right) }  }  \left(\rp{e_{\rm B}}{e_{\rm A}}\right)^2\left(\rp{-1+\sqrt{1-e^2_{\rm A}}}{ -1+\sqrt{1-e^2_{\rm B}} } \right)\right|\lb{mega}\eqf
If the modification of the gravitational potential (\ref{pot}), not modelled by Pitjeva in estimating $\Delta\dot\varpi$, accounts for what is unmodelled in the perihelia precessions, i.e. just for $\Delta\dot\varpi$, then (\ref{mega})  must be compatible with zero, within the errors.
\begin{table*}
\small
\caption{ A B denotes the pair of planets used; $\Pi=\Delta\dot\varpi_{\rm A}/\Delta\dot\varpi_{\rm B}$, $\mathcal{A}=(a_{\rm B}/a_{\rm A})^{1/2}$ and $F(e_{\rm A}, e_{\rm B})$ is given by \rfr{effe}. The perihelion extra-rates for the inner planets have been retrieved from \citep{Pit05b}; their errors are not the formal, statistical ones. The  perihelion extra-rates for the outer planets come from \citep{Pit06}; their formal errors have been re-scaled by a factor 10.  The uncertainties in the semimajor axes have been retrieved from Table IV of \citep{Pit05a}: they are the formal ones, but, as can be noted, their impact is negligible.  While $\Pi$ is always compatible with zero, this is definitely not the case for $\mathcal{A}$. The eccentricity function $F$ is always close to unity.
\label{tavola3}}

\begin{tabular}{@{}clll@{}}
\hline
A B & $\Pi$ & $\mathcal{A}$ & $F(e_{\rm A}, e_{\rm B})$\\
\tableline
 Mars Mercury & $-0.03\pm 0.2$ & $0.504\pm \mathcal{O}(10^{-12})$ & 1.008\\
 Mercury Jupiter & $-0.6\pm 4.1$ & $3.666\pm \mathcal{O}(10^{-9})$ &  0.989\\
 Earth Jupiter & $-0.03\pm 0.25$ & $2.281\pm \mathcal{O}(10^{-10})$ & 1.000\\
 Mars Jupiter & $ 0.02\pm 0.17$ & $1.847\pm \mathcal{O}(10^{-10})$ & 0.998\\
 Mercury Saturn & $ 0.004\pm 0.017$ & $4.963\pm \mathcal{O}(10^{-9})$ & 0.989\\
 Earth Saturn & $0.0002\pm 0.0011$ &   $ 3.088\pm \mathcal{O}(10^{-9})$ & 1.000\\
 Mars Saturn & $-0.0001\pm 0.0009$ & $2.501\pm \mathcal{O}(10^{-9})$ & 0.998\\
 Jupiter Saturn & $-0.006\pm 0.060$ & $1.353\pm \mathcal{O}(10^{-9})$ & 1.000\\
 Mercury Uranus & $-0.006\pm 0.152$ & $7.041\pm \mathcal{O}(10^{-8})$ & 0.989\\
 Earth Uranus & $ -0.0003\pm 0.0087$ & $ 4.380\pm \mathcal{O}(10^{-8})$ & 1.000\\
 Mars Uranus & $ 0.0002\pm 0.0048$ & $3.459\pm  \mathcal{O}(10^{-8})$ & 0.983\\
 Jupiter Uranus & $0.01\pm 0.31$ & $1.920\pm \mathcal{O}(10^{-8})$ & 0.999\\
\hline
\end{tabular}
\end{table*}
It must be noted that our approach is able to directly test the hypothesis that the proposed $1/r$ exotic force is not zero, irrespective of the magnitude of $\alpha$ because our prediction for the ratio of the extra-rates of perihelia, i.e. \eqi \rp{\mathcal{P}_{\rm A}}{\mathcal{P}_{\rm B}}= \sqrt{ \rp{ a_{\rm B}\left(1-e^2_{\rm A}\right) }{ a_{\rm A}\left(1-e^2_{\rm B}\right) }  }  \left(\rp{e_{\rm B}}{e_{\rm A}}\right)^2\left(\rp{-1+\sqrt{1-e^2_{\rm A}}}{ -1+\sqrt{1-e^2_{\rm B}} } \right),\eqf is just independent of $\alpha$ itself and is a specific function of the planets' orbital parameters $a$ and $e$. Of course, the ratio of the perihelion rates cannot be used, by definition, to test the zero hypothesis which, instead, can be checked by considering the apsidal precessions separately; indeed, in this case the prediction for the ratio of the perihelion precessions would be, by definition, $0/0$. Note that, being $\Delta\dot\varpi$ observational quantities, their ratio $\Pi=\Delta\dot\varpi_{\rm A}/\Delta\dot\varpi_{\rm B}$   is a well defined quantity, with an associated uncertainty $\delta\Pi$   which will be evaluated below.

 From \rfr{mega} and Table \ref{tavola3} it is possible to obtain
\begin{equation}\left\{\begin{array}{lll}
\Psi_{\rm MarMer} = 0.5\pm 0.2,\\\\
 \Psi_{\rm MerJup} = 4.2\pm 4.1,\\\\
 \Psi_{\rm EarJup} = 2.3\pm 0.2,\\\\
 \Psi_{\rm MarJup} = 1.8\pm 0.2,\\\\
 \Psi_{\rm MerSat} = 4.91\pm 0.02,\\\\
 \Psi_{\rm EarSat} = 3.090\pm 0.001,\\\\
  \Psi_{\rm MarSat} = 2.4984\pm 0.0008,\\\\
    \Psi_{\rm JupSat} = 1.36\pm 0.06,\\\\
    \Psi_{\rm MerUra} = 6.9\pm 0.1,\\\\
   \Psi_{\rm EarUra} = 4.383\pm 0.009,\\\\
    \Psi_{\rm MarUra} = 3.543\pm 0.005,\\\\
     \Psi_{\rm JupUra} = 1.9\pm 0.3.
      \lb{kaput}\end{array}\right.\end{equation}
The exact eccentricity-dependent factor
\eqi F(e_{\rm A},e_{\rm B}) = \sqrt{\rp{ 1-e^2_{\rm A} }{ 1-e^2_{\rm B} }  } \left(\rp{e_{\rm B}}{e_{\rm A}}\right)^2 \left(\rp{-1+\sqrt{1-e^2_{\rm A}}}{ -1+\sqrt{1-e^2_{\rm B}} }\right)\lb{effe}\eqf   of \rfr{mega}
is always close to unity, so that its impact on the results of \rfr{kaput} is negligible.
The errors in $\Psi_{\rm AB}$ due to $\delta a$ and $\delta\Delta\dot\varpi$ have been conservatively computed as
\begin{eqnarray}
\delta\Psi_{\rm AB} & \leq &
\left|\rp{\Delta\dot\varpi^{\rm A}}{\Delta\dot\varpi^{\rm B}}\right|\left(\rp{\delta\Delta\dot\varpi^{\rm A}}{|\Delta\dot\varpi^{\rm A}|}+
 \rp{\delta\Delta\dot\varpi^{\rm B}}{|\Delta\dot\varpi^{\rm B}|}\right) + \nonumber \\
  & & + \rp{1}{2}\left(\rp{a^{\rm B}}{a^{\rm A}}\right)^{\rp{3}{2}}\left(
\rp{\delta a^{\rm A}}{a^{\rm A}} + \rp{\delta a^{\rm B}}{a^{\rm B}} \right)F(e_{\rm A}; e_{\rm B});
 \end{eqnarray}
we did not optimistically summed the biased terms in a root-sum-square fashion because of the existing correlations among the  estimated extra-precessions, although they are low with a
maximum of about\footnote{L.I. thanks E.V. Pitjeva for such an information.} $20\%$ between Mercury and the Earth.
It turns out that by re-scaling the formal errors in the semimajor axes released by \citet{Pit05a} by a factor 10, 100 or more the results of \rfr{kaput} do not change because the major source of uncertainty is given by far by the errors in the perihelion precessions.  In regards to them let us note that the tightest constraints come from the pairs involving Saturn and Uranus\footnote{But not for the pairs A=Uranus/Saturn, B=Saturn/Uranus yielding values of $\Psi_{\rm AB}$ compatible with zero.} for which \citet{Pit05a} used only optical data, contrary to Jupiter\footnote{This is why a re-scaling of 10 of the formal error in its perihelion extra-precession should be adequate.}. Now, even by further re-scaling by a factor 10 the errors in their perihelion extra-rates, i.e. by a factor 100 their formal, statistical uncertainties, the results do not change (apart from A=Jupiter B=Uranus).  In the case of the pairs A=Earth, B=Uranus and A=Mars, B=Uranus even if the real error in the correction to the perihelion precession of Uranus was 1,000 times larger than the estimated formal one   the answer of \rfr{kaput} would still remain negative. For A=Mars and B=Saturn a re-scaling of even a factor 10,000 of the formal error in the perihelion extra-rate of Saturn would be possible without changing the situation.  In the case A=Earth, B=Saturn a re-scaling of 1,000 for the Saturn's perihelion extra-precession would not alter the result.

As a conclusion, the logarithmic correction to the Newtonian potential of \rfr{pot} is ruled out by the present-day observational determinations of the Solar System's planetary motions.
\section{The power-law corrections in the $f(R)$ extended theories of gravity}\lb{Capozzia}
In recent years there has been a lot of
interest in the so called  $f(R)$ theories of gravity \citep{sotfar08}. In them,
the gravitational Lagrangian
depends on an arbitrary analytic function $f$ of the Ricci scalar
curvature $R$ (see \citep{Capofranc07}  and references therein). These theories are also referred to as ``extended theories of gravity'',
since they naturally generalize, on a geometric ground,  GR, in the sense that when $f(R)=R$ the action reduces to the usual Einstein-Hilbert one, and Einstein's theory is obtained. It has been showed \citep{Capofranc07} that these theories    provide an alternative approach to solve, without the need of dark energy, some puzzles connected to the current cosmological observations and, furthermore, they can explain the dynamics of the rotation curves of the galaxies, without requiring dark matter. Indeed, for instance \citet{Capo07},  starting from $f(R)=f_0R^k$, obtained a power-law correction to the Newtonian gravitational potential of the form
\eqi V_{\beta}=-\rp{GM}{r}\left(\rp{r}{r_c}\right)^{\beta},\lb{capopot}\eqf (where $\beta$ is related to the exponent $k$ of the Ricci scalar $R$), and they  applied \rfr{capopot} to a sample
of 15 low surface brightness (LSB) galaxies with combined HI and H$\alpha$ measurements
of the rotation curve extending in the putative dark matter dominated region. They obtained
a very good agreement between the theoretical rotation curves and the data using only
stellar disk and interstellar gas when the slope $k$ of the gravity Lagrangian is set to the
value $k = 3.5$ (giving $\beta = 0.817$) obtained by fitting the SNeIa Hubble diagram with
the assumed power-law $f(R)$ model and no dark matter.

Here we wish to put on the test \rfr{capopot} in the Solar System with the approach previously examined.

First, let us note that an extra-radial acceleration
\eqi\bds A_{\beta} = \rp{(\beta -1)GM}{r_c^{\beta}}r^{\beta-2} \ \bds{\widehat{r}}\lb{accelcapo}\eqf can be obtained from \rfr{capopot}; let us now work out the secular precession of the pericentre of a test particle induced by \rfr{accelcapo} in the case $\beta-2<0$. By proceeding as in Section \ref{periprec} we get\footnote{See also \citep{Adk07}.}
\eqi\left<\dot\varpi\right>_{\beta}=\rp{(\beta-1)\sqrt{GM}}{2\pi r_c^{\beta}}a^{\beta-\rp{3}{2}}G(e; \beta),\lb{ge}\eqf
with
\eqi G(e; \beta)=-\rp{\sqrt{1-e^2}}{e}\int_0^{2\pi}\rp{\cos E-e}{(1-e\cos E)^{2-\beta}}dE.\lb{palle}\eqf   Since we are interested in taking the ratios of the perihelia, there is no need to exactly compute \rfr{palle}: the eccentricities of the Solar System planets are small and similar for all of them, so that we will reasonably assume that $G(e_{\rm A; \beta})/G(e_{\rm B}; \beta)\approx 1$.  Note that\footnote{The case $\beta=1$ is not relevant because it would yield a constant additional potential and no extra-force.} $\beta=0$ would yield a vanishing apsidal precession because $G(e;0)=0$. In fact, the case $\beta=0$ is compatible with all the estimated extra-rates of Table \ref{tavola1} and Table \ref{tavola2}; $\beta=0$, within the errors, is also the outcome  of different tests performed in the Solar System by \citet{Zak06}.

\subsection{Comparison with data}\lb{compa1}

According to  \citet{Capo07}, a value of $\beta=0.817$ in \rfr{capopot} gives   very good agreement between the theoretical rotation curves and the data, without need of dark matter. Again, if we consider a pair of planets A and B and take the ratio of their estimated extra-rates of perihelia,  by taking $\beta=0.817$, we may define \eqi \Xi_{\rm AB}=\left|\rp{\Delta\dot\varpi_{\rm A}}{ \Delta\dot\varpi_{\rm B} }-\left(\rp{a_{\rm B}}{a_{\rm A}}\right)^{0.683}\right|. \lb{xi}\eqf   Of course, such kind of  test could not be applied to the $\beta=0$ case since, in this case, we would have the meaningless prediction ${{\mathcal{P}}_{\rm A}}/{{\mathcal{P}}_{\rm B}}=0/0$.
If the modification of the gravitational potential of \rfr{capopot}  exists and is accounted for by the estimated corrections $\Delta\dot\varpi$
to the standard Newton-Einstein perihelion precessions, then the quantities $\Xi_{\rm AB}$  must be compatible with zero, within the errors. Instead, what we obtain, from Table \ref{tavola1}, Table \ref{tavola2}, Table \ref{tavola4} and \rfr{ge}, is
\begin{table*}
\small
\caption{ A B denotes the pair of planets used; $\Pi=\Delta\dot\varpi_{\rm A}/\Delta\dot\varpi_{\rm B}$, $\mathcal{B}=(a_{\rm B}/a_{\rm A})^{0.683}\ (\rm corresponding\ to\ \beta=0.817)$. The perihelion extra-rates for the inner planets have been retrieved from \citep{Pit05b}; their errors are not the formal, statistical ones. The  perihelion extra-rates for the outer planets come from \citep{Pit06}; their formal errors have been re-scaled by a factor 10.  The uncertainties in the semimajor axes have been retrieved from Table IV of \citep{Pit05a}: they are the formal ones, but, as can be noted, their impact is negligible.  While $\Pi$ is always compatible with zero, this is definitely not the case for $\mathcal{B}$.
\label{tavola4}}

\begin{tabular}{@{}clll@{}}

\hline

A B & $\Pi$ & $\mathcal{B}$ \\
\tableline
 Earth Mercury & $0.05\pm $ 0.19&   $0.523\pm \mathcal{O}(10^{-13})$ \\
 Mars Mercury & $-0.03\pm 0.2$ & $0.392\pm \mathcal{O}(10^{-13})$ \\
 Mercury Jupiter & $-0.6\pm 4.1$ & $5.898\pm \mathcal{O}(10^{-9})$ \\
 Earth Jupiter & $-0.03\pm 0.25$ & $3.084\pm \mathcal{O}(10^{-9})$ \\
 Mars Jupiter & $ 0.02\pm 0.17$ & $2.313\pm \mathcal{O}(10^{-10})$ \\
 Mercury Saturn & $ 0.004\pm 0.017$ & $8.921\pm \mathcal{O}(10^{-8})$ \\
 Earth Saturn & $0.0002\pm 0.0011$ &   $ 4.665\pm \mathcal{O}(10^{-9})$ \\
 Mars Saturn & $-0.0001\pm 0.0009$ & $3.499\pm \mathcal{O}(10^{-9})$  \\
 Jupiter Saturn & $-0.006\pm 0.060$ & $1.512\pm \mathcal{O}(10^{-9})$  \\
 Mercury Uranus & $-0.006\pm 0.152$ & $14.383\pm \mathcal{O}(10^{-8})$  \\
 Earth Uranus & $ -0.0003\pm 0.0087$ & $ 7.522\pm \mathcal{O}(10^{-8})$ \\
 Mars Uranus & $ 0.0002\pm 0.0048$ & $5.642\pm  \mathcal{O}(10^{-8})$  \\
 Jupiter Uranus & $0.01\pm 0.31$ & $2.438\pm \mathcal{O}(10^{-8})$  \\
\hline
\end{tabular}
\end{table*}

 \begin{equation}\left\{\begin{array}{lll}
\Xi_{\rm EarMer} = 0.5\pm 0.2,\\\\
\Xi_{\rm MarMer} = 0.4\pm 0.2,\\\\
\Xi_{\rm MerJup} = 6.5\pm 4.2,\\\\
\Xi_{\rm EarJup} = 3.1\pm 0.2,\\\\
\Xi_{\rm MarJup} = 2.3\pm 0.2,\\\\
\Xi_{\rm MerSat} = 8.92\pm 0.02,\\\\
\Xi_{\rm EarSat} = 4.666\pm 0.001,\\\\
\Xi_{\rm MarSat} = 3.4997\pm 0.0008,\\\\
\Xi_{\rm JupSat} = 1.52\pm 0.06,\\\\
\Xi_{\rm MerUra} = 14.4\pm 0.1,\\\\
\Xi_{\rm EarUra} = 7.523\pm 0.008,\\\\
\Xi_{\rm MarUra} =  5.642\pm 0.005,\\\\
\Xi_{\rm JupUra} = 2.4\pm 0.3.
\lb{kaputcapo}\end{array}\right.\end{equation}
It is remarkable to note that  even if the formal error in the Saturn's apsidal extra-precession was re-scaled by a factor 10,000 instead of 10, as done in this paper, the  pairs A=Earth B=Saturn and A=Mars B=Saturn would still rule out \rfr{ge}. A re-scaling by 1,000 of the Uranus' perihelion extra-precession would still be fatal, as shown by the pairs  A=Earth B=Uranus and A=Mars B=Uranus.

Thus, also the power-law correction to the Newtonian potential of \rfr{capopot} with $\beta=0.817$ is ruled out.
Criticisms to $R^k$ models of modified gravity were raised on different grounds by \citet{critica} who proposed more realistic models in \citep{pallee}.  Moreover, it is well-established now that DM shows also particle-like properties. In this respect, the proposal of $R^k$ gravity as DM (thanks to a change of the Newton potential) is not considered as a realistic one now. A more realistic DM candidate from $f(R)$ gravity was suggested by \citet{i_soliti}. They show that  not only a correction to the Newton potential appears, but also composite graviton degree of freedom shows particle-like behavior, as requested by DM data.

\section{Comments and conclusions}\lb{conc}

In this paper we have studied the secular
precession of the pericentre of a test particle in motion around a
central mass $M$ whose Newtonian gravitational potential  exhibits
a correction which has a logarithmic and power-law behavior. In order to put on the test the hypothesis that such extra-forces are not zero we devised a suitable test by taking into account the ratios (and not the extra-rates of the perihelia of each
planet at a time separately, or a linear combination of them, since their
uncertainties would fatally prevent to obtain any useful
constraints) of the
corrections to the secular precessions of the longitudes of perihelia estimated by E.V. Pitjeva for several pairs of planets in the Solar System.
The results obtained, resumed by \rfr{kaput} and \rfr{kaputcapo}, show that modifications of the Newtonian potentials like those examined in this paper are not compatible with the currently available  apsidal extra-precessions of the Solar System planets. Moreover, the hypothesis that the examined exotic force terms are zero, which cannot be tested by definition  with our approach, is  compatible with each perihelion extra-rate separately, in agreement with our results.
It must be noted that to give the hypothesized modifications of the Newtonian law the benefit of the doubt, and given that the formal errors in $a$ and $\Delta\dot\varpi$ of the outer planets are probably underestimates of the true uncertainties, we multiplied them by a factor ten or even more, as suggested to one of us (L.I.) by some leading experts in the ephemerides generation field like E.M. Standish, but the answers we obtained were still negative.  Another thing that should be pointed out is that, in principle, in assessing $\Psi_{\rm AB}$ and $\Xi_{\rm AB}$ one should have used $\Delta\dot\varpi^{\ast}=\Delta\dot\varpi-\delta\dot\varpi_{\rm canonical}$, where $\delta\dot\varpi_{\rm canonical}$  represents the mismodelled part of the modelled standard Newton/Einstein precessions. However, neglecting them did not affect our conclusions, as can be easily noted by looking at the $\Pi$ and ${\mathcal{A}}/{\mathcal{B}}$ columns in Table \ref{tavola3} and Table \ref{tavola4}. Indeed, the residual precessions due to the imperfect knowledge of, e.g., the solar quadrupole mass moment \citep{Pit05b} $\delta J_2/J_2\approx 10\%$ are of the order of $\leq 10^{-3}$ \asec \citep{Ior07c}, and even smaller are the mismodelled precessions due to other potential sources of errors like the asteroid ring or the Kuiper Belt objects \citep{IorPPN}.

However, caution is in order because, at present, no other teams of astronomers have estimated their own corrections to the Newtonian/Einsteinian  planetary perihelion rates, as it would be highly desirable.
If and when we will have, say, two independent determinations of the anomalous perihelion rate of a given planet $x=x_{\rm best}\pm \delta x$ and  $y=y_{\rm best}\pm \delta y$ we could see if they are compatible with each other and take their difference $|x_{\rm best}-y_{\rm best}|$ as representative of the real uncertainty affecting the apsidal extra-precession  of that planet.
Moreover, it would be interesting to see if for different sets of estimated corrections to the perihelion rates the figures for $\Pi$ in Table \ref{tavola3} and Table \ref{tavola4} change by an extent sufficient to alter the conclusions of  \rfr{kaput} and \rfr{kaputcapo}.

\section*{Acknowledgments}
M.L. Ruggiero acknowledges
financial support from the Italian Ministry of University and Research (MIUR) under the national program ``Cofin
2005'' - \textit{La pulsar doppia e oltre: verso una nuova era della ricerca sulle
pulsar}. \\


\end{document}